\documentclass[12pt]{article}

\usepackage{newtxtext,newtxmath}
\usepackage{amsmath}
\usepackage{physics}
\usepackage{graphicx}

\usepackage[letterpaper,margin=1in]{geometry}

\renewenvironment{abstract}
	{\quotation}
	{\endquotation}

\date{}

\makeatletter
\renewcommand{\fnum@figure}{\textbf{Figure \thefigure}}
\renewcommand{\fnum@table}{\textbf{Table \thetable}}
\makeatother

\usepackage{scicite}

\usepackage{url}

\usepackage[normalem]{ulem}

\def\scititle{
 Room-temperature magnon--phonon transduction in high-damping Co/Pt structures
}
\title{\bfseries \boldmath \scititle}

\author{
	Gauravkumar~Patel$^{1,\ast}$,
	Takuma~Sato$^{2}$,
	Maximilian~Frenzel$^{3}$,
	Prakriti~P.~Joshi$^{3}$,\and
	Ruslan~Salikhov$^{1}$,
	Ievgeniia Korniienko$^{4}$, 
	Dominik Legut$^{4,5}$,
	Olav~Hellwig$^{1,6}$,\and
    Sebastian~F.~Maehrlein$^{3,7,8}$,
	Kilian~Lenz$^{1}$,
	J\"urgen~Lindner$^{1,\ast}$\and
	\small$^{1}$Institute of Ion Beam Physics and Materials Research, Helmholtz-Zentrum Dresden--Rossendorf,\and
	\small Bautzner Landstra\ss e 400, 01328 Dresden, Germany.\and
	\small$^{2}$nextnano Lab, 22 chemin du Vieux Chêne, 38240 Meylan, France.\and
	\small$^{3}$Department of Physical Chemistry, Fritz Haber Institute of the Max Planck Society, 14195 Berlin, Germany.\and
	\small$^{4}$IT4Innovations, VSB - Technical University of Ostrava, 708 00 Ostrava, Czech Republic.\and
	\small$^{5}$Faculty of Mathematics and Physics, Charles University, 121 16 Prague, Czech Republic. \and
	\small$^{6}$Institute of Physics, Chemnitz University of Technology, 09107 Chemnitz, Germany.\and
	\small$^{7}$Institute of Radiation Physics, Helmholtz-Zentrum Dresden--Rossendorf,\and
	\small Bautzner Landstra\ss e 400, 01328 Dresden, Germany.\and
	\small$^{8}$Institute of Applied Physics, Dresden University of Technology, 01062 Dresden, Germany.\and
	\small$^\ast$Corresponding authors. Email: g.patel@hzdr.de, j.lindner@hzdr.de\and
}

\begin{document} 

% Insert the title and author list
\maketitle

% Abstract, in bold
% There are strict length limits, and not all formats have abstracts.
% Consult the journal instructions to authors for details.
% Do not cite any references in the abstract.
\begin{abstract} \bfseries \boldmath
% Start with one or two sentences of background
Quantum communication and information processing strongly benefit from the coupling between different quasi-particles, offering complementary advantages. Magnetoelastic materials inherently allow for direct coupling between magnetization dynamics and quantized lattice vibrations, called phonons. Near the ferromagnetic resonances, phonons may thus trade energy and angular momentum with uniformly precessing magnetization, called magnons, and enable transduction of information from magnetic to phononic modes, thereby paving the way for long-range transport of magnetic information without the need of magnetic material. Here, we employ tailored magnetic-nonmagnetic heterostructures, which simultaneously act as cavities for standing shear waves, to bring selective phonons and magnons into resonance. These Co films with Pt seed layers show extended linewidth and reduced amplitude of the phonon-resonant FMR lines, providing a hallmark of energy and angular momentum exchange. Complementarily, by theoretical modeling and ultra-fast coherent phonon spectroscopy, we identify the responsible transverse acoustic phonons as standing shear waves in the combined Co and Pt structure. We find a high crystal quality in conjunction with a large magnetoelastic coupling constant as a prerequisite for efficient magnon--phonon coupling of this type. Such resonant enhancement of magnon--phonon coupling in CMOS-compatible material provides an ideal material platform for future quantum transducers.
\end{abstract}

% The first paragraph of any Science paper does NOT have a heading
% Nor is it indented
\section*{Introduction}
\label{se:intro}

Hybridization of quantized excitations is a key element to integrate distinct physical systems with complementary characteristics for quantum communication and information processing \cite{Li_2020,Lachance_Quirion_2019,Kurizki_2015,Xiang_2013}.
In particular, magnetoelastic coupling links magnetization precession to mechanical lattice vibrations, forming hybrid magnon–phonon states that facilitate coherent exchange of energy and angular momentum \cite{An_2020,Seavey_1965,Bommel_1959,Muller_2024}. 
This hybrid platform provides a promising route to efficient transduction between spin and mechanical angular momentum for future memory devices and sensors \cite{Hann_2019,Wallucks_2020,Bandyopadhyay_2021,Lachance_Quirion_2019}.
Extensive studies using low-damping yttrium iron garnets (YIG) have demonstrated strong magnon–phonon coupling and long-range transfer of angular momentum via phonons, with propagation distances that can exceed characteristic electron spin-diffusion and magnon-propagation lengths \cite{An_2023,An_2020,Ruckriegel_2020,Li_2021,Sato_2021,Schlitz_2022,An_2022}.
Moreover, the broken time-reversal symmetry of magnetization dynamics, together with the conservation of crystal angular momentum~\cite{Minakova_2026}, enable the conversion of spin angular momentum to chiral or axial phonons \cite{Juraschek_2025}, giving rise to directional phonon transport and new spin–mechanical phenomena that can be employed for nonreciprocal devices and spin-to-mechanics transduction \cite{Muller_2024,Holanda_2018,Verba_2019,Shah_2020}.
Alongside quantum transduction and spin-vibrational conversion, magnetization-driven excitation of lattice vibrations, termed phonon pumping, has emerged as an important mechanism for spintronic functionality \cite{Streib_2018,Sato_2021,Rezende_2021,Peria_2022,Zhang_2020,Lenk_2011,Chumak_2015}.
Spintronic devices typically employ metallic ferromagnet/heavy-metal (FM/HM) heterostructures, which combine low-cost fabrication and CMOS compatibility with efficient current-induced magnetization control via spin-transfer and spin–orbit torques \cite{Krivorotov_2005,Lenk_2011,Chumak_2015,Manchon_2019,Song_2021}. 
Recent theoretical work further predicts that, in addition to spin angular momentum (spin currents), orbital angular momentum (orbital currents) can be converted into phonon angular momentum, establishing a tripartite coupling among spin, orbital, and lattice degrees of freedom \cite{Han_2025}.
Despite these advances, phonon pumping at FM/HM metallic interfaces---where strong interfacial spin–orbit coupling, broken inversion symmetry, and enhanced magnetoelastic interactions coexist---remains largely unexplored.
Studying phonon emission and angular-momentum transfer across such interfaces is essential both for testing theoretical predictions and to enable practical spintronic transducers that harness spin-orbital-phonon conversion. 
Therefore, systematic experimental and theoretical investigations of phonon pumping at FM/HM interfaces are crucial for advancing studies of orbital currents and for the development of next-generation spintronic and hybrid technologies.

In this article, efficient phonon pumping at the prototypical spintronic Co/Pt interface is demonstrated despite large magnetic and acoustic damping at room temperature, contrary to CoFe films~\cite{Müller_2024}, where high acoustic losses suppresses the phonon pumping at room temperature. 
Transverse acoustic (TA) phonons are launched by the magnetization dynamics in the Co layer that propagate through the Pt layer, while the silicon substrate used in the system acts as a phonon sink. 
Using conventional ferromagnetic resonance (FMR) spectroscopy, signatures of anticrossings between the uniform (Kittel) magnon mode and Co$\vert$Pt-combined-thickness standing phonon modes are observed. 
The concurrent mode hybridization and abrupt increase in FMR linewidth at these anticrossings provides clear evidence of magnon–phonon cross-talk and the opening of an additional relaxation channel for angular momentum. 
The resonance frequency of the standing phonon depends on the Pt thickness, indicating that FMR-pumped phonons can traverse the Pt layer despite its acoustic mismatch with Co \cite{Sato_2021}. 
Furthermore, we demonstrate that the magnon--phonon transduction efficiency is tunable via the sample design: optimization of the structural quality in the Co layer enhances the effect, whereas no phonon-pumping signature is observed for Co/Ta interfaces, where Co films grow in a polycrystalline manner \cite{Patel_2023}. 
These findings imply that in ferromagnetic films with perpendicular magnetic anisotropy, phonon pumping constitutes a significant relaxation channel, enhancing magnon–phonon transduction and opening prospects for efficient applications in technologically relevant, non‑epitaxial films.

 \section*{Results}
\label{se:fmr}

\begin{figure}[t!]
\centering
\includegraphics[width=0.8\textwidth]{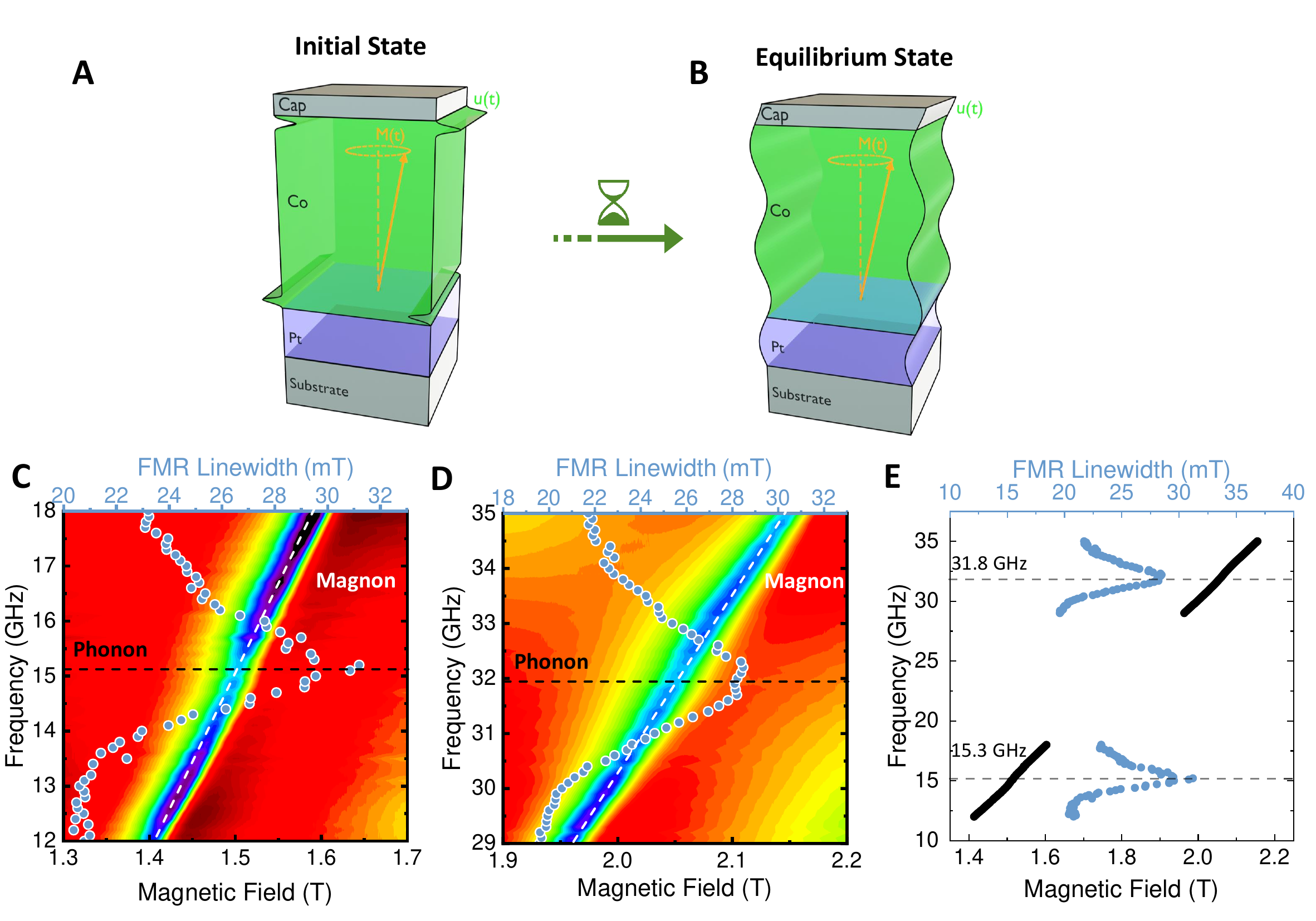}
\caption{Schematic illustrating the generation of phonons. $u(t)$ is induced by magnon–phonon coupling at the Co interfaces. (\textbf{A}) depicts the initial transient state, while (\textbf{B}) shows the subsequent formation of a transverse acoustic (TA) standing phonon mode during the shear-wave equilibrium state. (\textbf{C}) and (\textbf{D}) are intensity maps of the amplitude of the high resolution FMR spectra for the 35-nm thick Co film. The FMR linewidth is plotted as blue circles with matching frequency axis. The intensity map depicts anti-crossings between the magnon and phonon modes that consequently broaden the FMR linewidths at those frequencies. (\textbf{E}) The FMR resonance field (black) and FMR linewidth $\Delta {H_\mathrm{pp}}$ (blue) are plotted together on matching frequency axis. The peaks in the linewidth occur at frequencies where there is a small distortion in the high resolution $f(H)$, i.e.\ at $15.3~\mathrm{GHz}$ and $\mathrm{31.8~\mathrm{GHz}}$, respectively.}
\label{fig:FMRFinescan}
\end{figure}

Thin Co films with thicknesses between 4 and 110~nm were grown by magnetron sputtering on a 20‑nm Pt seed layer deposited on thermally oxidized Si(001).
The Pt seed provides a template for a hexagonal texture, which is responsible for large magnetic anisotropy \cite{Patel_2023} and an ordered medium for phonons to propagate.
A schematic of this stack with resulting shear waves $u(t)$ is shown in Figs. \ref{fig:FMRFinescan}\textbf{A-B}. Figures~\ref{fig:FMRFinescan}\textbf{C-D} show intensity maps of FMR spectra of a 35-nm-thick Co film as a function of external magnetic field in out-of-plane (OOP) geometry.
In addition, the extracted FMR linewidth is shown with matching frequency axis. 
In OOP geometry, the uniform magnon (Kittel) mode increases linearly with frequency as visible by the white dashed guide lines. 
We observe weak anti-crossings around 15~GHz and 32~GHz in conjunction with an increase in FMR linewidth around those frequencies as shown in Fig.~\ref{fig:FMRFinescan}\textbf{C-D}. 
Figure \ref{fig:FMRFinescan}\textbf{E} shows the FMR resonance field and the FMR linewidth ($\Delta {H_\mathrm{pp}}$) extracted by fitting the high resolution FMR measurement in out-of-plane (OOP) geometry of a 35-nm-thick Co film.
As a result of two weak anti-crossings, the resonance field dependence shows two slight deviations from linearity and the resulting peaks in the linewidth are centered at 15.3~GHz and 31.8~GHz, respectively.
The enhanced FMR linewidth indicates an enhanced energy dissipation of the magnetization precession at those frequencies.
As two‑magnon–scattering–induced broadening is minimal in the OOP configuration \cite{Landeros_2008} and can thus be ruled out, we assign the strong linewidth increase to signatures of magnon--phonon hybridization, where phonons absorb energy and angular momentum from the precessing magnetization.
Since magnetic and elastic losses in the metallic system are higher than in insulators like YIG \cite{An_2020}, instead of an anti-crossing gap, only a deviation of the linear $f(H)$ dependence is observed. 
The maximum linewidth, which lies at the center of the anti-crossing, indicates a maximum energy transfer between the interacting modes. 
In other words, the frequency of the linewidth maximum likely serves as a proxy for the phonon eigenmodes of the bilayer sample and will be investigated in the following section.

\begin{figure}[t]
\centering
\includegraphics[width=1\textwidth]{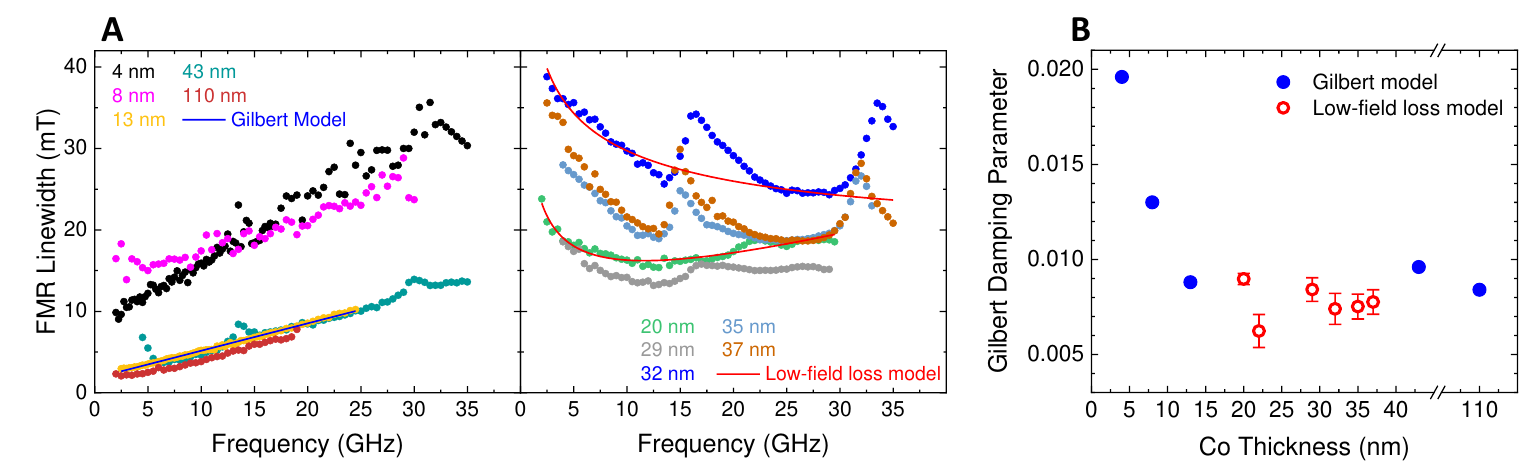}
\caption{(\textbf{A}) Frequency-dependence of the peak-to-peak FMR linewidth $\Delta H_\mathrm{pp}$ in out-of-plane (OOP) geometry for the Co thin films. The linewidth dependence for the Co thickness is split in two plots for visual clarity. The blue and red curves show fits using the Gilbert model Eq.~(\ref{eq:Gilbertmodel}) only and including the low-field loss model Eq.~(\ref{eq:lowfieldloss}), respectively. (\textbf{B}) Gilbert damping constants extracted from both models.
}
\label{fig:LWPeak}
\end{figure}

To investigate the hybridization, the Co thickness was varied from 4~nm to 110~nm. 
As seen in Fig.~\ref{fig:FMRFinescan}, the hybridization results in a faint anticrossing but strongly enhances the FMR linewidth, serving as a robust metric for tracking hybridization over the Co thickness range. 
In Fig.~\ref{fig:LWPeak}, the FMR linewidth dependence is shown for all investigated samples. 
For better visibility of the data, the sample set is divided into two panels. 
It is observed that not all the thicknesses show local maxima in the linewidth dependence.
For Co thicknesses of 4~nm to 13~nm, and 110~nm, the linewidth shows a linear dependence without any peaks in the linewidth, and the Gilbert damping parameter $\alpha$ can be extracted from fits using Gilbert's model \cite{Heinrich_2005}:
\\
\begin{equation}
\label{eq:Gilbertmodel}
    \Delta {H_\mathrm{pp}} = \Delta {H_\mathrm{pp}^{0}}  + \frac{2}{\sqrt{3}} \frac{2\pi\alpha}{\gamma}\,f,
\end{equation}
\\
\noindent where $\gamma$ is the gyromagnetic ratio and $\Delta {H_\mathrm{pp}^{0}}$ is some inhomogeneous linewidth broadening as a result of magnetic inhomogeneity in the system. The extracted values of $\alpha$ are plotted in Fig.~\ref{fig:LWPeak}\textbf{B} as solid circles. 
The other samples with Co thicknesses of 20~nm to 43~nm show a nonlinear and nonmonotonic frequency dependence of the linewidth. 
A significant decrease of the linewidth at lower frequencies followed by one or two linewidth peaks at higher frequencies is observed.
Hence, the determination of the Gilbert damping parameter $\alpha$ requires modeling of the linewidth increase towards lower frequency. 
At low frequencies, when the corresponding resonance field is also low, the magnetization does not align with the external field but with the internal anisotropy fields.
This results in a broadening of the linewidth, often known as low-field loss \cite{Nembach_2011,Schloemann_1992}. 
The influence of this mechanism drops by $1/f^n$ upon increasing the frequency \cite{Nembach_2011}. 
Hence, Eq.~(\ref{eq:Gilbertmodel}) is extended to:
\\
\begin{equation}
\label{eq:lowfieldloss}
    \Delta {H_\mathrm{pp}} = \Delta {H_\mathrm{pp}^{0}} + \Delta H_\mathrm{pp}^\mathrm{low} \frac{1}{f^n} + \frac{2}{\sqrt{3}} \frac{2\pi\alpha}{\gamma}\,f,
\end{equation}
\\
\noindent and termed low-field loss model in this article.

Excluding the frequency range of the peaks in the linewidth dependence, the exemplary fits are shown as red lines in Fig.~\ref{fig:LWPeak}\textbf{A}. 
The Gilbert damping constants $\alpha$ extracted using this model are plotted in Fig.~\ref{fig:LWPeak}\textbf{B} with open red circles. The $\alpha$ from both models roughly follow the same trend as expected for magnetic thin films \cite{Barati_2014,Sharma_2022}.

\begin{figure}[t]
\centering
\includegraphics[width=1\textwidth]{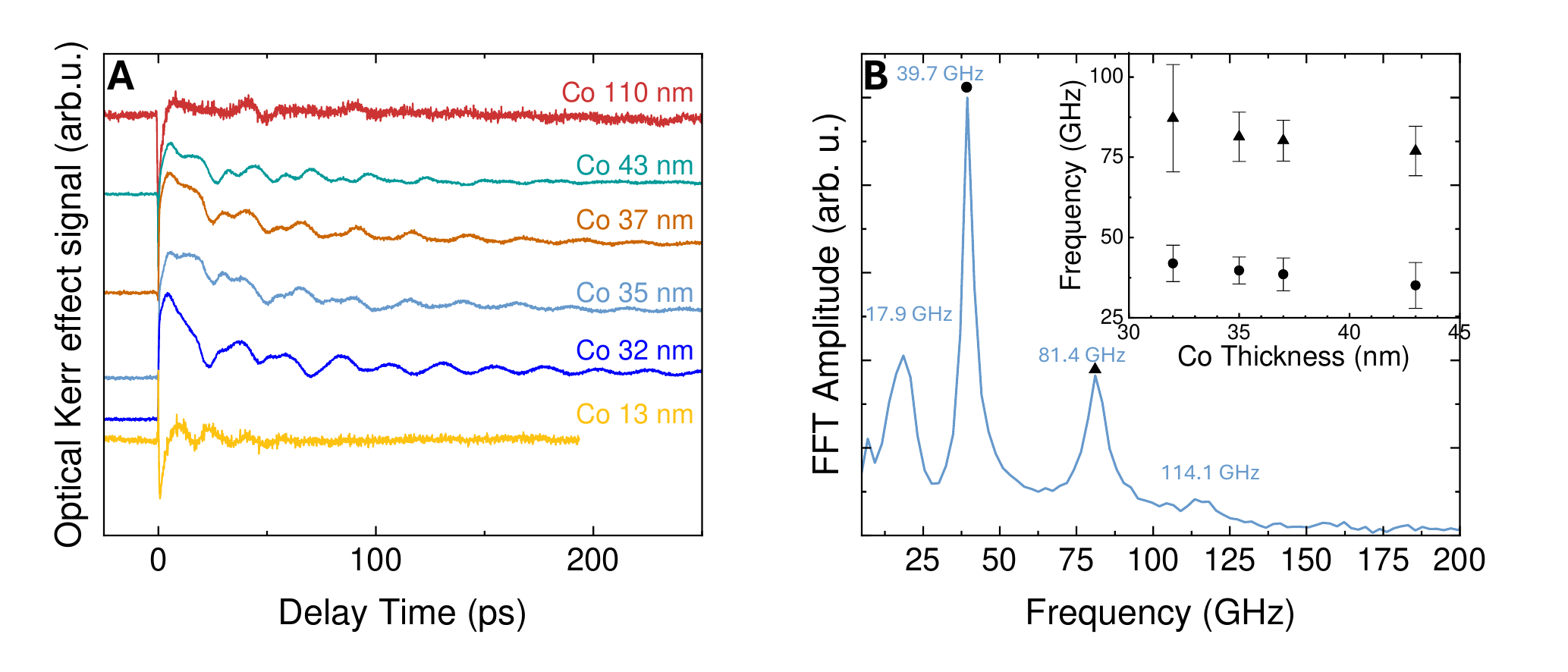}
\caption{(\textbf{A}) Optical Kerr effect (OKE) signal from various Co films. The oscillatory signal is dominated by coherent acoustic wave packets launched at the surface and then traveling back and forth through the heterostructure. (\textbf{B}) In frequency space, i.e.\ the FFT of the background subtracted signal of \textbf{A}, this corresponds to standing acoustic phonons given by the resonance condition of the Co+Pt hybrid cavity, shown here exemplarily for 35~nm Co. The inset shows the thickness dependence of two dominant phonon resonances indicated by circles and triangles.}
\label{fig:pump-probe}
\end{figure}

Now, we focus on peaks in the linewidth dependence as a result of hybridization. 
The peaks in Fig.~\ref{fig:LWPeak}\textbf{A}, shift towards lower frequency upon increasing the Co layer thickness. 
Given such an inverse relationship with thickness and magnetoelastic properties of Co, the frequency peaks may be identified as standing phonon modes, excited as a result of dynamic magnetoelastic coupling from the uniformly precessing magnetization at FMR. 

In the following, an independent investigation of the acoustic resonances in the form of standing phonon modes was carried out by the optical Kerr effect (OKE) measurements based on time-resolved pump-probe spectroscopy.
For this measurement, an ultrashort optical pump pulse excites a strain wave, which travels as a phonon wave packet, periodically modulating the linear susceptibility at the probe pulse wavelength, leading to a transient anisotropic reflectivity change as a function of pump-probe delay. \cite{De_Silvestri_1985,Frenzel_2023,Maehrlein_2021}. 
Figure \ref{fig:pump-probe}\textbf{A} shows a measured oscillatory signal as a function of the delay time between pump and probe pulse. To extract the oscillations from the measured signal, the time-dependent incoherent background due to heating effects was removed using a biexponential fit. The fast-Fourier transform (FFT) of the extracted oscillations was then used to determine the frequencies of the corresponding phonon resonances. Figure \ref{fig:pump-probe}\textbf{B} shows the FFT of the 35-nm-thick Co film only. Given the accuracy of the pump-probe spectroscopy in the limited time window, the first and second phonon peak at 17.9 $\pm$ 9.5~GHz and 39.7 $\pm$ 4.3~GHz roughly match with the FMR linewidth peak positions, which were 15.3 $\pm$ 0.85~GHz and 31.8 $\pm$ 1.2~GHz for the 35-nm-thick Co film, respectively. Furthermore, the pump-probe measurement reveals even higher order standing phonon modes at 81.4~GHz and 114.1~GHz. Their FMR detection was not possible due to the limited frequency and magnetic field range of the FMR setup. As shown in the inset, upon increasing the Co thickness, the frequency of the phonon mode decreases, which matches well with the thickness dependent trend observed from the FMR measurements in Fig.~\ref{fig:LWPeak}\textbf{A}.

\begin{figure}[t!]
\centering
\includegraphics[width=1\textwidth]{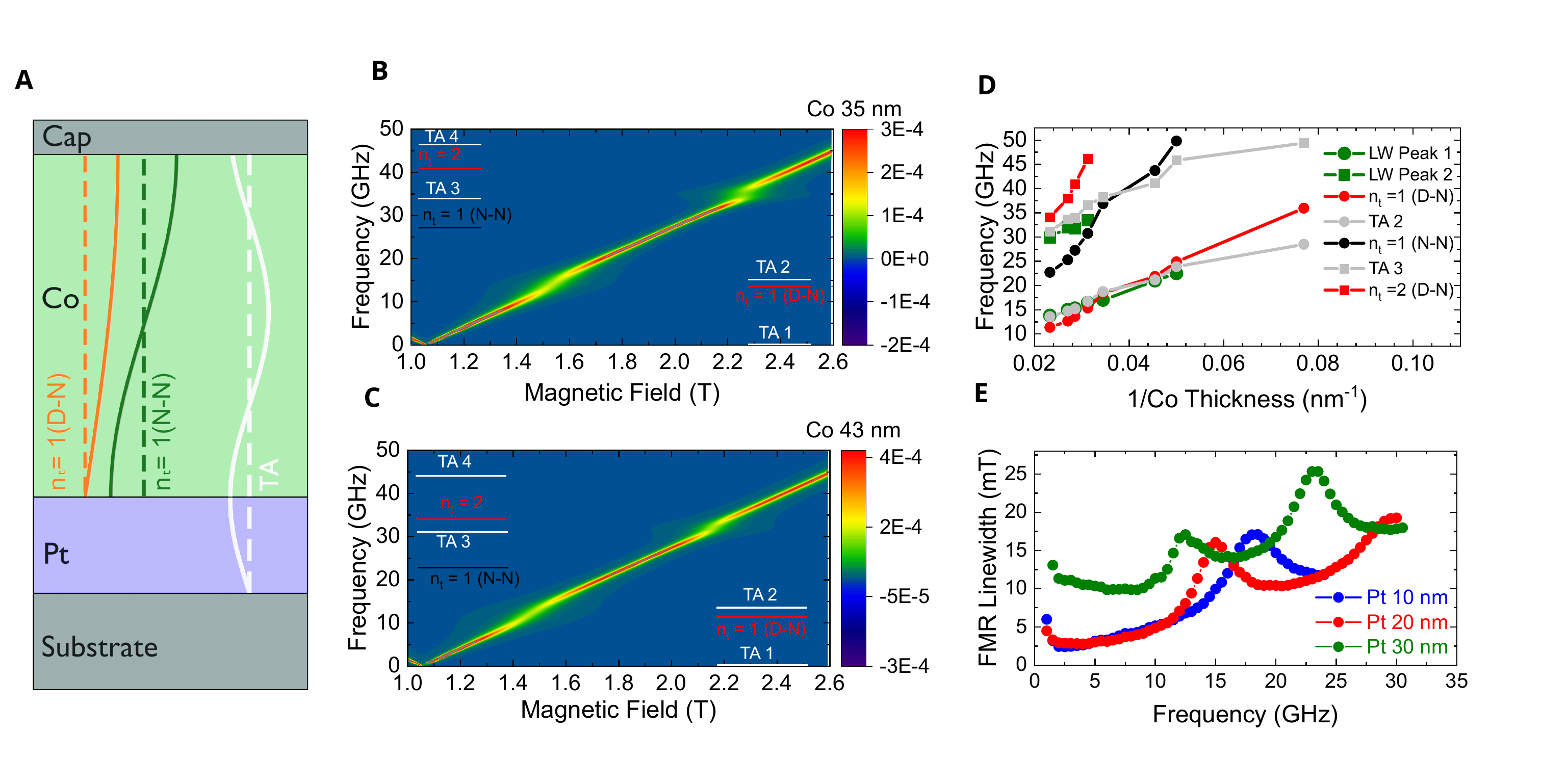}
\caption{(\textbf{A}) Schematic representation of the various standing phonon modes. 
Intensity maps of the calculated Im[$\chi_\mathrm{tot}(f)$] from Eq.~(\ref{eq:chi-total}) of the (\textbf{B}) 35-nm-thick and (\textbf{C}) 43-nm-thick Co films. The frequencies of all phonon modes are marked by small lines and labels. TA stands for transversal acoustic mode. The free and fixed boundary conditions at the interfaces for the phonon vibration are labeled as `N' and `D', respectively. For example, $n_t=1$(D-N) denotes a first-order phonon mode with fixed boundary at the bottom interface and free at the top. (\textbf{D}) The thickness dependence of the experimental linewidth (LW) peak positions (green), calculated frequencies at which phonon pumping is efficient due to the stress matching~\cite{Sato_2021} ($n_t$, red and black), and calculated TA phonon modes (gray). The circular symbols indicates the phonon mode closer to the first anti-crossing in the FMR $f(H)$ dependence and square symbols indicate phonon modes closer to the second anti-crossing. (\textbf{E}) FMR linewidth dependence of the Pt thickness series at fixed Co thickness (35~nm).}
\label{fig:FMR Absorption-Theory}
\end{figure}

To get a detailed understanding about these phonon modes, we employ a theoretical model based on the magnon--phonon coupling \cite{Sato_2021}. 
At FMR, the uniformly precessing magnetization coherently induces a dynamic magnetoelastic stress, which exerts force on the lattice only at the top and bottom interfaces of the magnetic material as schematically shown in Fig.~\ref{fig:FMRFinescan}\textbf{A} and excites transverse phonons, which propagate along the thickness\cite{Sato_2021,Seavey_1965}. 
At the steady state, the transverse phonons can form various standing waves along the thickness as schematically depicted in Fig.~\ref{fig:FMR Absorption-Theory}\textbf{A} as curves. 
In OOP geometry, the magnetization cannot hybridize with the longitudinal phonons and is excluded from discussion \cite{Sato_2021,Kittel_1958,Kobayashi_1973}.
In Ref.~\cite{Peria_2022} it was assumed that the standing phonon was only generated inside the magnetic layer. 
This means that the higher-order phonon modes should be an integer multiple of the fundamental mode frequency, which depends on the phonon group velocity, thickness, and interface boundary conditions \cite{Photinos_2017}. 
We observe from the FMR linewidth peak positions in Fig.~\ref{fig:FMRFinescan} and from the FFT peak positions from Fig.~\ref{fig:pump-probe} that the second mode does not occur at twice the fundamental frequency but slightly higher. This motivates to look deeper into the origin of the observed phonon modes.

Therefore, the phonon group velocity, thickness, and boundary conditions for the Pt layer were also taken into account, because the phonon wave packet, after reflection from the Pt/substrate interface, is retransmitted into the Co layer again. 
Thereby, it interferes with itself and forms a standing phonon mode across the \emph{combined} thickness of Co and Pt. 
The elastic standing modes, i.e.\ solutions to Eq.~(\ref{eq:EOM}), only form in the bilayer when the wavelength matches an integer multiple of the thickness.
The bilayer therefore acts as a Fabry-P\'erot cavity for acoustic phonons. 
The resonance condition for forming the standing wave in the combined thickness of Co and Pt, which is a transverse acoustic (TA) mode, is \cite{Sato_2021}:
 \\
\begin{equation}
\label{eq:stadingphononcondition}
    \sin \tilde{k}_td\cos k_tL_1 + \frac{\rho v_t}{\tilde{\rho}\tilde{v}_t}\cos\tilde{k}_td\cos k_tL_1 = 0,
\end{equation}
\\
\noindent where $d$ is the thickness of the magnetic (Co) layer, $L_1$ is the thickness of the Pt layer, $\tilde{k}_t$ is the phonon wave vector in Co, $\tilde{\rho}$ is the mass density of Co and $\tilde{v}_t$ is the transverse speed of sound in Co. The quantities without tilde hold for Pt. 
The term $\rho v_t$ is known as the acoustic impedance. 
The acoustic impedance mismatch, i.e., $\rho v_t/\tilde{\rho}\tilde{v}_t = 2.06$ in Co/Pt, is responsible for the higher harmonics being non-integer multiples of the fundamental mode, as already observed in Fig.~\ref{fig:LWPeak}\textbf{A} and \ref{fig:pump-probe}\textbf{B}, representing the acoustic analogue of a hybrid optical cavity \cite{Spencer_2025}.
Figure \ref{fig:FMR Absorption-Theory}\textbf{A} schematically depicts three different standing-mode profiles in the Co-Pt bilayer, 
which can couple to the magnons. 
The red and black phonon mode profiles fulfill the Neumann boundary condition at least at one of the Co surfaces, and get strongly amplified by the surface magnetoelastic stress generated by the Kittel mode (stress-matching condition \cite{Sato_2021}). 
The conceptual TA eigenmode profile depicted in white may or may not have nodes at the Co surfaces, but it is the central channel of phonon pumping due to its large density of states. 
In Fig.~\ref{fig:FMR Absorption-Theory}, `D-N' and `N-N' refer to the combinations of fixed-end (Dirichlet, D) and free-end (Neumann, N) boundary conditions at the two Co interfaces, respectively.

Further, the absorbed FMR intensity Im$\left(\chi_\mathrm{tot}\left(f\right)\right)$ was calculated by our model detailed in the supplementary material, and is plotted in Figs.~\ref{fig:FMR Absorption-Theory}\textbf{B} and \textbf{C} for the 35-nm-thick and 43-nm-thick films, respectively. 
The following parameters were used: $\tilde{\rho} = 8930 \,\mathrm{kg}/\mathrm{m^{3}}$, $\tilde{v}_t = 1960 \,\mathrm{m/s}$, $\Tilde{\eta} = 1\,\mathrm{GHz}$~\cite{High-eta}, $\rho = 21450 \,\mathrm{kg}/\mathrm{m^{3}}$, $v_t = 1680 \,\mathrm{m/ s}$, $\eta = 10\,\mathrm{GHz}$~\cite{High-eta}, $K_2 = 0.51 \,\mathrm{MJ/m^3}$~\cite{Patel_2023}, and $\alpha = 0.0058$. 
Here, the speed of sound in the Co layer ($\Tilde{v}_t$) is assumed smaller than its bulk value, because sputter-grown Co contains predominantly columnar grains and stacking faults, which reduce the speed of sound in Co thin films, as discussed in the supplementary material.

Instead of an expected linear dependence between magnetic field and frequency in OOP geometry, here, multiple anti-crossing signatures are observed---directly reflecting the hybrid magnon--phonon state. 
As already discussed from Fig.~\ref{fig:FMRFinescan}, the faint anti-crossing as a result of magnon--phonon coupling leads to the peaks in the FMR linewidth, presenting clear hallmarks of the weak coupling regime. 
Now, to identify which standing phonon modes interact with the magnon, all the expected phonon modes in the displayed frequency range are marked in Fig.~\ref{fig:FMR Absorption-Theory}\textbf{B}--\textbf{C}.
It can be seen that around both anti-crossings, more than one mode ($n_t =1(\mathrm{D-N})$ and TA2, $n_t =2(\mathrm{D-N})$ and TA3) could hybridize with magnons as they are very close to each other at the anti-crossing frequency. 
To identify the relevant modes, the thickness dependence of the resonance frequency of all four modes is plotted together with the linewidth peak positions (green symbols) from the FMR measurement in Fig.~\ref{fig:FMR Absorption-Theory}\textbf{D}. 
From this it is evident that both modes, TA-2 and TA-3, match very well with the measured peak frequencies in the broader thickness range, while $n_t$=1 and $n_t$=2 show a slightly different thickness dependence. 
Moreover, the calculation also predicts higher-order TA modes at 80.04~GHz and 114.20 GHz for the 35-nm-thick film, which match very well with the higher-order modes observed in the coherent phonon response in Fig.~\ref{fig:pump-probe}\textbf{B}.

\section*{Discussion}
\label{se:discussion}

Hence, we conclude that the standing modes of the Co/Pt bilayer, denoted as TA modes in Fig.~\ref{fig:FMR Absorption-Theory}\textbf{B}--\textbf{C}, hybridize with the magnons and provide a channel for converting angular momentum and energy from the spintronic to the phononic system, representing magnon--phonon transduction. 
It is to be noted that the magnon--phonon transduction was achieved regardless of larger acoustic damping due to the Landau-Rumer mechanism at room temperature \cite{Müller_2024,de_Klerk_1965,1965187} in sputter-deposited Co/Pt. Thus, making it attractive for CMOS-compatible transducers.
To our knowledge, this is the first experimental demonstration of a correlation between the standing TA phonon-mode spectrum and its thickness dependence at room temperature in sputter deposited films. Furthermore, to experimentally cross-check that the TA mode exists in a combined thickness of Co and Pt, another sample series was fabricated, where the Pt thickness was varied from 10~nm to 30~nm, while keeping the Co thickness fixed at 35~nm. 
This sample series had to be fabricated in a different sputter system, but with similar growth parameters. 
The FMR linewidth dependence of these samples is shown in Fig.~\ref{fig:FMR Absorption-Theory}\textbf{E}. 
As shown in Fig.~\ref{fig:LWPeak}\textbf{A} and Fig.~\ref{fig:FMR Absorption-Theory}\textbf{E}, the linewidth-peak positions of both 35-nm-thick Co on 20-nm-thick Pt films deposited in the two different sputter systems match. This proves the reproducibility of the samples and the phenomena found. Most importantly, increasing the Pt thickness, the linewidth peak position shifts towards lower frequencies. 
This confirms that the phonon pumping excites standing waves across the Co/Pt bilayer. This result is particularly striking when compared to conventional YIG/GGG systems. 
Since the acoustic impedance of YIG and GGG matches, TA eigenmodes of the bilayer are equidistant in frequency, allowing for thickness tuning that aligns the ladder of the (N-N) modes with that of the TA modes. 
This facilitates observation of phonon-pumping-induced anticrossings \cite{An_2020}. 
Our experiments, however, show that phonon pumping can be tailored in bilayers by acoustic impedance mismatch.

Finally, after understanding the origin of the FMR linewidth peak and the underlying phonon mode, we address the optimization of magnon-phonon transduction efficiency by understanding the peak amplitude in Fig.~\ref{fig:LWPeak}. 
For thicknesses of 20 and 29~nm, the peak amplitudes are smaller than for the 32-, 35-, and 37-nm-thick Co films. For the 43-nm-thick Co film, the peak amplitude decreases again. 
The amplitude of the peak in the linewidth dependence—a direct measure of transduction strength—shows that phonon pumping is most efficient at intermediate thicknesses.
In a previous structural characterization of these films \cite{Patel_2023}, it was found that at lower thickness, the Co films are strained and at higher thickness, due to fcc stacking fault multiplication, a significant amount of fcc Co grains introduce inhomogeneity along the film thickness. 
This results in structural defects at lower and higher thicknesses.
Consequently, the magnetocrystalline anisotropy, which closely depends on the crystal structure, was found to be smaller than for samples of intermediate thickness, where Co achieves an optimal crystal structure with fewer stacking faults.
Hence, the better crystal quality, which reduces the scattering of phonons, is responsible for the efficient magnon-phonon transduction in hcp Co thin films. 
Complementary to the amplitude of the peak in linewidth dependence, this is also evident in our pump-probe experiments. 
While the coherent lattice vibrations persisted up to 400~ps at the intermediate thicknesses, in thinner (15~nm) and thicker (43 and 110~nm) films, the phonon decay is faster due to multiple phonon scattering at structural defects such as strain and stacking faults in the Co films. 
Hence, in thinner and thicker samples, the large elastic damping blurs the TA mode spectra, thereby reducing the efficiency of the magnon--phonon conversion. 
Consequently, the rapid, defect-driven decay of these phonons impedes the formation of a robust acoustic standing wave. 
Without a strongly defined standing wave to facilitate the coherent transfer of energy and momentum from the uniform precession, the FMR linewidth peak is significantly reduced in the 43-nm-thick Co film, and entirely suppressed in the 110-nm-thick sample. 
The efficient magnon-phonon transduction in the Co/Pt structure highlights the viability of these structures as robust acoustic sources. 
Given the large phonon propagation length, the phonon pumping enables the long-range transduction of magnonic information carried by the angular momentum of the TA phonons in the phononic system without the need for a continuous magnetic system \cite{Minakova_2026}.
Traditionally, such transduction required highly complex, pristine single crystals (e.g.,\ YIG/GGG) with ultra-low magnetic and acoustic damping. 
However, our demonstration of phonon pumping in sputter-deposited Co/Pt systems offers highly scalable, CMOS-compatible pathways towards long-range quantum transduction.

\section*{Conclusion}
The realization of coherent spin-mechanical transduction in scalable architectures is a critical milestone for integrating spintronic and phononic information transfer into next-generation quantum hybrid technologies. In this work, we have demonstrated that highly efficient magnon-phonon hybridization can be achieved at room temperature in prototypical, high-damping Co/Pt structures. By directly correlating out-of-plane FMR with time-resolved pump-probe measurements, we unambiguously observed resonant phonon pumping and massive linewidth enhancement driven by the hybridization of the uniform magnon mode with transverse acoustic standing phonons.
Based on theoretical calculations and experimental results, we have shown that the \emph{combined} thickness of Co and Pt forms an acoustic cavity. 
The magnon-phonon transduction efficiency is heavily dependent on interfacial structural quality, maximizing at an intermediate Co thickness that provides the structural "sweet spot" to host long-lived TA phonons.
This coherent transduction enables long-range transport of magnonic information by the phonons without a magnetic channel at room temperature. 
Our work demonstrates that practical device engineering for magnon-phonon transduction is no longer restricted to ideal, low-loss insulators.
By realizing robust phonon pumping at a prototypical FM/HM interface, characterized by strong spin–orbit coupling, opens vital experimental pathways to explore the recently predicted tripartite coupling among spin, orbital, and lattice degrees of freedom. 
Ultimately, the ability to harness this spin-to-mechanics conversion in standard thin-film structures offers a highly scalable, technologically relevant route toward integrating phononic information transfer into next-generation memory, sensor, and quantum hybrid technologies.

\subsection*{Materials and Methods}
\textbf{Fabrication of Co thin-films}\\
The Co films were fabricated at room temperature by DC magnetron sputtering in an ultrahigh vacuum system (AJA International ATC2200) with a base pressure of about $10^{-7}$~mbar. 
The Ar pressure was kept at $4\times10^{-7}$~mbar during the deposition and the Ar flow rate was fixed to 25~sccm. 
Si(001) wafers with a 100-nm-thick thermally oxidized $\mathrm{SiO_2}$ surface layer were used as substrates.
A 1.5-nm-thick Ta layer was deposited onto the substrates for better adhesion. 
Further, a 20-nm-thick Pt seed layer was deposited to promote growth of the hexagonal close packed crystal structure of Co. 
The thickness of the Co films was varied from 4~nm to 110~nm. All samples were finally capped by 3~nm of Pt to prevent oxidation.\\

\noindent\textbf{Vector-Network-Analyzer Ferromagnetic Resonance}\\
The broadband FMR measurements were performed on a home-built spectrometer using a Keysight N5527A VNA connected to a coplanar waveguide with the sample mounted flip-chip on it. 
By sweeping the external magnetic field at constant frequency, the microwave transmission parameter $\abs{S_{21}}$ was recorded as the FMR signal. 
Sweeping through FMR, $\abs{S_{21}}$ follows a Lorentzian lineshape whose center field is the resonance field. 
The linewidth values are given as peak-to-peak linewidth $\Delta H_\mathrm{pp}$ as used in field-modulated FMR for easy comparison. 
The FMR linewidth is a collective measure of all the loss mechanism and inhomogeneity present in the system. \\

\noindent\textbf{Ultrafast Optical Kerr Effect Measurement}\\
In the ultrafast optical Kerr effect (OKE), an intense laser pulse (pump) induces a transient anisotropic refractive index change (transient birefringence), which causes a change of the polarization state of a time-delayed optical probe pulse \cite{Maehrlein_2021}. 
To implement this, 400-nm-wavelength laser pulses (at 1~kHz repetition rate, $\sim$1~mJ/cm$^2$, and $\sim$270~fs pulse duration) were generated by second harmonic generation in a BBO crystal, using 800 nm laser pulses from a Ti:Sapphire amplifier. 
The probe laser pulses were supplied from a synchronized Coherent Vitara-T Ti:sapphire oscillator, delivering pulses at 80~MHz repetition rate with a center wavelength of 800~nm and a pulse duration of 20~fs.
These probe pulses were collinearly aligned with and temporally delayed relative to the optical pump pulses. 
The induced phase difference was recorded using a balanced detection scheme consisting of a half-waveplate and a Wollaston prism, which spatially separates the perpendicularly polarized components of the probe beam. 
The intensities of the two separated beams were measured by a pair of photodiodes and the difference signal corresponds to the anisotropic change of the refractive index. 
This technique enables the time-resolved detection of the coherent Raman-active phonons, which are driven by the optical pump pulse via impulsive stimulated Raman scattering \cite{De_Silvestri_1985}. 
The collective Raman-active phonon motion modulates the linear susceptibility at the probe pulse wavelength; thus, imprinting the coherent phonon dynamics on the measured difference signal.\\

\noindent\textbf{Theoretical Model}\\
The model considers an FM/NM bilayer with a surface normal pointing along the $z$-axis. Under the application of an external magnetic field, the magnetization precession at resonance frequency $\omega=2\pi f$ exerts a coherent dynamic stress due to the magnetoelastic coupling to the underlying system. This generates phonons with frequency $\omega$ \cite{Sato_2021,Seavey_1965}. At the interfaces, the phonons are partially reflected as well as transmitted along the thickness into the adjacent layers. The equation of motion for the lattice vibration can be written as follows; 
\begin{align}
\label{eq:EOM}
    \frac{\partial^2u_\alpha}{\partial z^2} + \frac{\omega^2}{\Tilde{v}_t^2}\left(1+i\frac{\Tilde{\eta_\mathrm{el}}}{\omega}\right)u_\alpha = 0  &\qquad \text{in FM} \\ \notag
    \frac{\partial^2u_\alpha}{\partial z^2} + \frac{\omega^2}{v_t^2}\left(1+i\frac{\eta_\mathrm{el}}{\omega}\right)u_\alpha = 0 &\qquad \text{in NM}, 
\end{align}

\noindent where the equation with tilde is for the ferromagnetic (FM) layer and without tilde is for the normal metal (NM) layers. $v_t$ is the transverse speed of sound and $\eta_\mathrm{el}$ is the elastic damping of the phonons. In this article, the NM consists of a 20-nm-thick Pt layer (thickness $L_1$) and the Si substrate (thickness $L_2$), which act as phonon sinks.
% Instead of developing a trilayer model, the bilayer model of Ref.~\cite{Sato_2021} is employed, and the presence of the Si substrate (phonon sink) was mimicked by a large elastic damping in the Pt layer.

Conversely, the lattice vibrations generate a small strain induced tickle field ($\Omega_{\mathrm{me}}^\mathrm{ij}$) on the magnetization as a result of the magnetoelastic coupling. 
Hence, the modified LLG equation can be written as follows~\cite{Sato_2021}:  

 \begin{align}
 \label{eq:LLG}
    \begin{pmatrix}
    m_x\\
    m_y
    \end{pmatrix} (\omega) = \chi(\omega,\theta_\mathrm{m})
    \left[\begin{pmatrix}
    h_x\\
    h_y
    \end{pmatrix}-\frac{1}{\gamma\mu_0}\begin{pmatrix}
    \Omega'^{13}_\mathrm{me} \\
    \Omega'^{23}_\mathrm{me}
    \end{pmatrix}\right](\omega), 
\end{align}

\noindent where $\chi(\omega,\theta_\mathrm{m})$ is the high-frequency susceptibility. 

In general, the LLG equation and the elastic equation of motion (EOM) constitute a fully coupled system that requires a self-consistent solution due to the dynamic back-action between the lattice displacement $u_\alpha$ and the magnetization $\mathbf{m}$. 
However, to calculate the FMR spectrum, we restrict our model to one iteration between the two equations. 
We treat the driven Kittel mode as an independent source term, determine $u_\alpha$ as a function of the magnetization vector $\mathbf{m}$ by solving the elastic EOM, and plug them into Eq.~(\ref{eq:LLG}). 
This renormalizes the susceptibility tensor as:

\begin{align}
\label{eq:chi-total}
    \begin{pmatrix}
    m_x\\
    m_y
    \end{pmatrix} (\omega) = \chi_{tot}(\omega,\theta_\mathrm{m})
    \begin{pmatrix}
    h_x\\
    h_y
    \end{pmatrix}(\omega).
\end{align}

\noindent The imaginary part of the renormalized susceptibility Im[$\chi_{tot}(\omega,\theta_\mathrm{m})]$ gives the FMR absorption spectrum as a function of frequency and external magnetic field, which also includes the influence of the magnetoelastic coupling on the FMR.
It is shown in Fig.~\ref{fig:FMR Absorption-Theory} as an intensity map.
\\

\noindent\textbf{Transveral phonon modes ($\tilde{v}_t$)---sound speed dependence in Co}\\
The speed of sound of the transverse phonon modes ($\tilde{v_t}$) in hcp and fcc structures is related to the elastic constants $C_{44}$  and $C'=\frac{C_{11}-C_{12}}{2}$ as $\tilde{v_t}=\sqrt{\frac{C_{44}}{\rho}}$ and $\tilde{v_t}=\sqrt{\frac{C'}{\rho}}$, respectively.
In addition, for the fcc structure, there is one more shear mode for propagation along the [111] direction, i.e., a transverse mode combining all cubic $C_{ij}$ with the speed $\tilde{v_t}=\sqrt{\frac{C_{11}-C_{12}+C_{44}}{3\rho}}$\cite{2001xi}. 
As the latter one gives numbers that are slightly higher than the other two let us comment on the shear modes common in both hcp and fcc structures. 
To investigate the role of grains on the speed of sound for transversal modes in hcp and fcc structures of Co, a molecular dynamics (MD) simulations using the \textsc{Lammps} software \cite{thompson2022lammps} and modified embedded atom method potential \cite{SHARIFI2025113595} is used. 
For fcc and hcp Co in all simulations, the following experimental lattice constants were used: $a_{lat}=3.545~\Angstrom$ \cite{WANG201911} and $a_{lat}= 2.507~\Angstrom$ with $c_{lat}=4.070~\Angstrom$, respectively \cite{Kittel}. The elastic constants $C_{ij}$ in GPa for single crystalline hcp and fcc structures are summarized in Table \ref{Table_cij}.

\begin{table}[bt]
\begin{center}
    \caption{\label{Table_cij}Elastic constants $C_{ij}$ for hcp and fcc structures of Co calculated at $T=0$~K and compared to the experimental values at $T=298$~K \cite{Cij_exp, Cij_exp2}. In order to compare the fcc to hcp structure the tensor of the $C_{ij}$ of fcc Co was rotated such that the principal c-axis is set along [111] and marked by `rot.'.}

    \begin{tabular}{ccccccc}
      \hline\hline$C_{ij}$ (GPa)   & $C_{11}$ & $C_{12}$ & $C_{13}$ & $C_{33}$ & $C_{44}$ & $C'$ \\
      \hline
     fcc Co & 269 & 149 && & 101 & \phantom{1}60\\ 
     fcc Co (rot.)  & 310 & 135 & 122 & 324 & 74 & \phantom{1}87\\
     hcp Co & 303 & 138 & 126 & 322 & 79& \phantom{1}83\\ 
    
      \hline
      experiment\cite{Cij_exp} & 307 & 165 & 103 & 358 & 75.5& \phantom{1}71\\  
      experiment\cite{Cij_exp2} & 320 & 66 & 102 & 374 & 82 & 127\\     
      \hline\hline
    \end{tabular}
    
    \end{center}
\end{table}

 In order to prepare the polycrystalline structures with an increasing number of grains, the software \textsc{Atomsk} \cite{HIREL2015212} and its built-in Voronoi tessellation technique, which allows structure tuning and the application of periodic boundaries, was used. The estimations of effective elastic constants in the MD simulations were based on deforming the simulation box in one of six directions (at zero temperature) and measuring the change in the stress tensor \cite{Clavier22112017}.
 
%-----------------
\begin{figure}[ht]
\centering
\includegraphics[width=0.6\columnwidth ,angle=0]{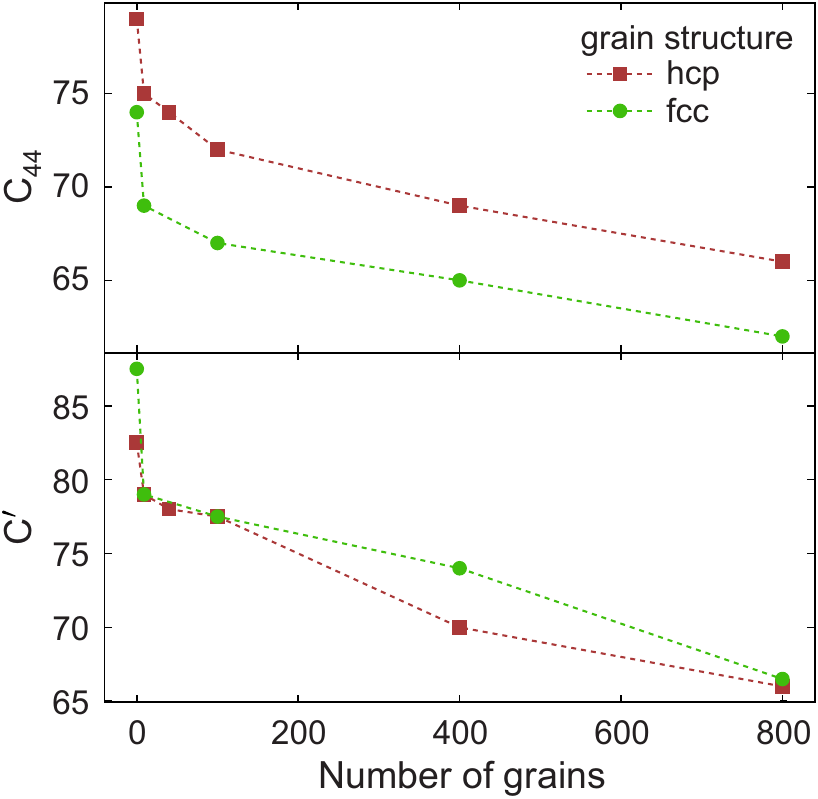}
\caption{Effective elastic constants $C_{44}$ and $C'$ as a function of number of columnar grains for hcp and fcc Co. For polycrystalline grains of fcc Co a rotated tensor of elastic constants was used for compatibility and comparison with polycrystalline hcp Co, i.e., the principal c-axis was along [111].}
%For hcp Co polycrystal the volume is $500\times 500 \times 40$ \AA$^3$ and fcc Co polycrystal the volume is $500\times 500 \times 42$ \AA$^3$.  }
\label{fig:fcccol-9}
\end{figure}
%-----------------
From Table \ref{Table_cij} one can see that $C'$ is greatly reduced by almost 30\% in fcc structures in contrast to hcp Co. 
However, in the columnar grains the orientation of the fcc phase is mainly along the [111] direction. Hence, the $C_{ij}$ are of similar magnitude [see the values for fcc Co (rot.) in Table \ref{Table_cij}]. 
Next, the effective $C_{ij}$ are determined for polycrystalline hcp material (sample size $500\times 500 \times 40~\Angstrom^3$) and fcc material ($500\times 500 \times 42~\Angstrom^3$) ranging from 9 to 800 columnar grains, respectively. 
The results are given in Fig.~\ref{fig:fcccol-9}. 
They clearly show that an increase in the number of grains in the system leads to a decrease in the effective elastic constants $C_{44}$ and $C'$, which in turn reduce the velocity of transverse waves. 
For the largest grain number used in the simulations (i.e.\ 800 grains), one can see that $C'$ is similar for both fcc and hcp structure amounting to 66--67~GPa. 
The trend of the reduction of $C_{44}$ as the number of grains increases is similar for hcp and fcc Co. 
However, still a smaller value is obtained for the fcc Co phase $C_{44}\approx 62$~GPa. It should also be noted that the density of polycrystalline structures differs from the density of a single crystal, but this difference is not significant. 
Therefore, for the 800 grains the speed of sound of the shear (transverse) modes is $\tilde{v_t} \approx$ 2778 (2681)~m/s from $C_{44}$ and $\tilde{v_t} \approx$ 2762 (2773)~m/s from $C'$ for hcp (fcc), respectively. 
Thus, the analysis qualitatively shows a tendency for the speed of sound to decrease due to the presence of grains in polycrystals. 
However, in real material, the crystallographic orientations in the grains are not purely random (for example, in Co, 4 types of twin boundaries can occur \cite{Hakamada2012}), the same as formation of the secondary tensile twins inside the primary compressive twins \cite{Wang2017} and stacking faults \cite{Patel_2023} might be presented which in turn may affect the quantitative characteristics of the effect.

\section*{Acknowledgments}
G.P. and R.S. acknowledge support from the German Research Foundation (DFG) Grant No. 566522216. D.L. and I.K. acknowledge computational time by the project e-INFRA CZ (ID:90254) by the Ministry of Education, Youth and Sports of the Czech Republic and support by project no. 22-35410K by the Czech Science Foundation. S.F.M. acknowledges support and funding from the Deutsche Forschungsgemeinschaft (DFG,
grant No. 469405347) for his Emmy Noether group ``Circular Phononics”. The authors thank T.\ Naumann for his experimental and technical support.

%%%%%%%%%%%%%%%% REFERENCES %%%%%%%%%%%%%%%

\clearpage % Clear all remaining figures and tables then start a new page

% The list of references goes after the main text and before the acknowledgements
% When preparing an initial submission, we recommend you use BibTeX, like this:
%

%%%%%%%%%%%%%%%% ACKNOWLEDGEMENTS %%%%%%%%%%%%%%%

\bibliography{LW_Paper} % for a file named science_template.bib
\bibliographystyle{sciencemag}

\end{document}